\documentclass[conference]{IEEEtran}
\IEEEoverridecommandlockouts
\usepackage{cite}
\usepackage{amsmath,amssymb,amsfonts}
\usepackage[ruled,vlined,linesnumbered]{algorithm2e}
\usepackage{graphicx}
\usepackage{textcomp}
\usepackage{xcolor}
\usepackage[hidelinks]{hyperref}
\usepackage{balance}

\makeatletter
\let\MYcaption\@makecaption
\makeatother

\makeatletter
\newcommand{\linebreakand}{%
  \end{@IEEEauthorhalign}
  \hfill\mbox{}\par
  \mbox{}\hfill\begin{@IEEEauthorhalign}
}
\makeatother

\usepackage[font=footnotesize]{subcaption}

\makeatletter
\let\@makecaption\MYcaption
\makeatother

\newtheorem{definition}{Definition}

\begin{document}

\title{A Parallel and Distributed Rust Library for Core Decomposition on Large Graphs\\
\thanks{This work is supported by Spoke 1 and Spoke 6 of the ICSC National Research Centre for High Performance Computing, Big Data and Quantum Computing within the NextGenerationEU program (Project Code: PNRR CN00000013).}
}

\author{\IEEEauthorblockN{1\textsuperscript{st} Davide Rucci}
\IEEEauthorblockA{\textit{ISTI-CNR} \\
Pisa, Italy \\
davide.rucci@isti.cnr.it}
\and
\IEEEauthorblockN{2\textsuperscript{nd} Sebastian Parfeniuc}
\IEEEauthorblockA{\textit{University of Pisa} \\
Pisa, Italy \\
s.parfeniuc@studenti.unipi.it}
\and
\IEEEauthorblockN{3\textsuperscript{rd} Matteo Mordacchini}
\IEEEauthorblockA{\textit{IIT-CNR} \\
Pisa, Italy \\
matteo.mordacchini@iit.cnr.it}
\linebreakand 
\IEEEauthorblockN{4\textsuperscript{th} Emanuele Carlini}
\IEEEauthorblockA{\textit{ISTI-CNR} \\
Pisa, Italy \\
emanuele.carlini@isti.cnr.it}
\and
\IEEEauthorblockN{5\textsuperscript{th} Alfredo Cuzzocrea}
\IEEEauthorblockA{\textit{University of Calabria} \\
Rende, Italy \\
alfredo.cuzzocrea@unical.it}
\and
\IEEEauthorblockN{6\textsuperscript{th} Patrizio Dazzi}
\IEEEauthorblockA{\textit{University of Pisa} \\
Pisa, Italy \\
patrizio.dazzi@unipi.it} 
}

\maketitle

\begin{abstract}

In this paper, we investigate the parallelization of $k$-core decomposition, a method used in graph analysis to identify cohesive substructures and assess node centrality. Although efficient sequential algorithms exist for this task, the scale of modern networks requires faster, multicore-ready approaches.
To this end, we adapt a distributed $k$-core algorithm originally proposed by Montresor et al. to shared-memory systems and implement it in Rust, leveraging the language’s strengths in concurrency and memory safety. We developed three progressively optimized versions: SequentialK as a baseline, ParallelK introducing multi-threaded message passing, and FastK further reducing synchronization overhead.
Extensive experiments on real-world datasets, including road networks, web graphs, and social networks, show that FastK consistently outperforms both SequentialK and ParallelK, as well as a reference Python implementation available in the NetworkX library. Results indicate up to an 11× speedup on 16 threads and execution times up to two orders of magnitude faster than the Python implementation.

\end{abstract}


\section{Introduction}

Graph-structured data have become a central abstraction across diverse application domains, from social and communication networks to biological interaction graphs, transportation systems, and the web itself~\cite{jiang2022graph,mordacchini2025decentralized,hetzel2021graph,mordacchini2025ICDCS,xue2025data}. In recent years, graph-based modeling has extended well beyond traditional network science, becoming a unifying framework for reasoning about interactions, dependencies, and optimization in heterogeneous systems. Graph representations have been proposed for centralized~\cite{agouti2022graph,lin2020guardian} and decentralized social platforms~\cite{guidi2019towards}, mobile communication networks~\cite{wang2018d2d,chen2014towards,mordacchini2017social}, vehicular ad hoc systems~\cite{santos2021temporal}, resource allocation~\cite{ferrucci2024decentralized} and load distribution in IoT/edge~\cite{wu2021internet,CARLINI2026108167,ferrucci2021latency} and fog computing~\cite{hong2019resource}. In the context of cloud infrastructures~\cite{zolghadri2024resource} and federated learning~\cite{li2024exploiting,akhtarshenas2024federated,kavalionak2021impact}, graphs capture data flow, task placement, and collaboration patterns among distributed agents. Similarly, graph abstractions are increasingly employed in cybersecurity, recommendation systems~\cite{gennaro2007mroute}, biological networks, and optimization of large-scale computing environments. This proliferation underscores the versatility of graph-based solutions as a foundation for both theoretical analysis and practical system design.

Given this wide range of applications, identifying key structural features of graphs is essential for both understanding and optimizing system behavior. Among the available graph analysis tools, the~\emph{k-core decomposition}~\cite{SEIDMAN1983269} stands out for its simplicity, interpretability, and analytical power. It provides a hierarchical peeling of the network, identifying progressively denser subgraphs that can reveal communities, influential actors, and structural weaknesses. The \emph{coreness} of a node represents the highest $k$ such that the node belongs to the $k$th core, which can be computed directly through standard linear-time algorithms~\cite{k_core_survey2020} (see also Section~\ref{sec:sota}). The coreness metric captures a network's community structure~\cite{Malvestio2020, santofortunato, Rucci:2025Asonam,BAE2014549} and serves as a meaningful node importance score~\cite{kumar2020identifying, wu2015core}, complementing measures such as degree and $h$-index. Its sensitivity to structural anomalies also makes it valuable for detecting outliers~\cite{faloutsos_corescope, Rucci:2024SAC}.
%

Despite the existence of efficient sequential algorithms, e.g. most notably the linear-time approach of Batagelj and Zaversnik~\cite{batagelj2003omalgorithmcoresdecomposition}, the rapid growth of real-world graphs in both size and density poses new computational challenges~\cite{sakr2021future}. Modern social or web graphs routinely include hundreds of millions of vertices and billions of edges, stressing the limits of single-core memory bandwidth and cache capacity. As data volumes continue to expand, exploiting multicore and many-core architectures becomes essential to preserve interactivity and scalability in graph analytics workflows.
This transition, however, is not straightforward: irregular memory access patterns, data dependencies, and load imbalance make graph algorithms notoriously difficult to parallelize effectively. A variety of parallel and distributed $k$-core algorithms have been proposed to address these challenges. Shared-memory systems typically rely on fine-grained parallel peeling, dynamic work distribution, or frontier-based processing, while distributed systems adopt message-passing protocols that exchange local degree estimates among neighboring vertices. Among the latter, the decentralized algorithm by Montresor et al.~\cite{montresor} offers an elegant, asynchronous formulation particularly suited for large-scale and failure-tolerant environments. Yet, adapting such distributed designs to shared-memory architectures raises non-trivial issues related to synchronization cost, cache coherence, and contention on shared data structures. Achieving high performance while preserving correctness and determinism requires careful engineering of communication patterns and memory layouts.

In this paper, we introduce a Rust-based parallel library for $k$-core decomposition, which leverages ideas from both distributed protocols and shared-memory execution. Rust provides a unique combination of low-level control and strong compile-time safety guarantees: its ownership model eliminates data races, while zero-cost abstractions and fine-grained concurrency primitives make it suitable for high-performance scientific computing. 
Ultimately, we design a shared-memory adaptation that maintains the decentralized spirit of message-passing while minimizing synchronization and communication overhead. 

We introduce three progressively optimized implementations. \texttt{SequentialK} serves as a correctness baseline and performance reference. \texttt{ParallelK} explores multi-threaded execution using message queues and synchronization barriers, leveraging both high-level thread-pool abstractions and native threads. Finally, \texttt{FastK} refines this design with cache-aware data structures, global shared states, and selective message propagation to reduce contention and improve convergence speed.

Through extensive experiments on large real-world graphs, including road networks, social networks, and web graphs, we show that \texttt{FastK} consistently achieves substantial gains over both our baselines and widely used Python implementations such as NetworkX~\cite{NetworkX}, delivering up to 11× speedup on 16 cores and execution times two orders of magnitude faster. Beyond performance, our results demonstrate the practical viability of Rust for graph analytics: it combines efficiency and memory safety without sacrificing low-level control. The proposed library contributes reusable building blocks for parallel and distributed graph algorithms, paving the way toward safer, high-performance data-intensive computation.

To present these results, the remainder of this paper is organized as follows. Section~\ref{sec:definition} provides the basic definitions and background about the $k$-core decomposition. Section~\ref{sec:sota} frames our work with the relevant scientific literature.  Section~\ref{sec:ParallelK} details parallel Rust implementations and their design choices. Section~\ref{sec:FastK} introduces \texttt{FastK} and its optimizations along with synchronization and activation strategies. Section~\ref{sec:exp} presents the experimental methodology, datasets, and results; Section~\ref{sec:conc} draws concluding remarks.

\section{Background and Related Work}~\label{sec:definition}\label{sec:sota}
The concept of the $k$-core, first introduced by Seidman~\cite{SEIDMAN1983269}, provides a fundamental tool for analyzing the structure of complex networks. Intuitively, a $k$-core is a maximal subgraph in which every vertex has at least $k$ neighbors within the subgraph. Formally:

\begin{definition}[$k$-core~\cite{SEIDMAN1983269}]
For any positive integer $k$, the $k$-core of a graph $G=(V,E)$ is the inclusion-maximal subset $K \subseteq V$ such that the induced subgraph $G[K]$ contains only vertices of degree at least $k$.
\end{definition}

The $k$-core decomposition of a graph consists of computing all possible $k$-cores for $k = 1, 2, \dots, k_{\max}$, where $k_{\max}$ is the largest integer for which a non-empty $k$-core exists. This decomposition induces a nested hierarchy of subgraphs, where each $(k+1)$-core is contained within the $k$-core, revealing progressively denser regions of the network. Vertices can be assigned a coreness value, defined as the largest $k$ for which the vertex belongs to the corresponding $k$-core. 

Coreness provides a simple yet informative centrality measure, often used to identify influential or structurally important nodes in networks.
Several properties make $k$-cores particularly useful for network analysis. First, the hierarchical structure obtained through peeling allows for the detection of cohesive subgroups or communities without requiring global optimization or parameter tuning. Second, the $k$-core can highlight resilient structures, as vertices in higher-order cores are typically better connected and more robust to vertex removal. Third, the computational simplicity of $k$-core decomposition makes it suitable for very large networks: the standard sequential algorithm by Batagelj and Zaversnik~\cite{batagelj2003omalgorithmcoresdecomposition} achieves linear time complexity with respect to the number of edges, $O(|E|)$.

Beyond theoretical interest, $k$-core decomposition has been widely applied in practice. In social networks, it is used to identify communities and influential users; in biological networks, it highlights essential proteins or interactions; in web and citation networks, it assists in ranking pages or authors by structural importance. Furthermore, $k$-core-based methods have been adopted in optimization problems for distributed systems, traffic flow analysis, and resource allocation, as the hierarchical subgraph structure often correlates with functional or operational significance.

Computing $k$-cores is conceptually straightforward but poses computational challenges for large-scale and dense graphs, particularly when parallel or distributed execution is desired. The iterative nature of the peeling process, combined with irregular degree distributions typical of real-world networks, creates load imbalance and irregular memory access patterns. These challenges motivate the development of efficient parallel algorithms, which aim to retain the correctness of the sequential decomposition while exploiting modern multi-core and many-core architectures.
In summary, the $k$-core decomposition provides a powerful lens to analyze graph structure, offering both a mathematically rigorous definition and practical tools for identifying dense and resilient substructures. Its hierarchical nature, ease of computation, and applicability across diverse domains make it a central technique in modern network science and graph analytics.

\subsection{Related Work}
The core decomposition and maintenance problem has been studied in the literature for a long time, as it is a polynomial-time solvable problem that provides much information about any network.
The centralized, state-of-the-art, algorithm widely adopted is due to Batagelj and Zaversnik~\cite{batagelj2003omalgorithmcoresdecomposition}.
It is a recursive, $O(|V|+|E|)$ algorithm that removes from the graph the node with minimum degree, which will become their coreness value.
From the theoretical point of view, this algorithm is optimal as it basically traverses the entire graph.
With the steady rise of the size of nowadays graphs, many parallel algorithms have been designed throughout the years, in order to boost computation efficiency.
\emph{ParK}, proposed by Kabir et al. \cite{Park_2014}, is a two-phase algorithm that works on \emph{levels}, groups of nodes such that their coreness is exactly $l$, the level.
ParK was later improved and further optimized by PKC \cite{PKC}.

Shun and Blelloch propose Ligra~\cite{Ligra2013}, a shared-memory parallel graph framework that provides a concise vertex- and edge-mapping abstraction (VERTEXMAP, EDGEMAP) for efficiently expressing traversal-based algorithms such as BFS, PageRank, and connected components. Ligra's main contribution, i.e., a hybrid sparse/dense frontier representation that dynamically adapts to the active vertex set, enables work-efficient\footnote{A parallel algorithm is \emph{work-efficient} if it performs asymptotically the same total amount of work as the best known sequential algorithm. For example, a work-efficient $k$-core algorithm runs in $O(|V| + |E|)$ total work, matching the optimal sequential bound.} graph computation in $O(|V| + |E|)$ total time. This design later served as the foundation for work-efficient k-core decomposition algorithms, which build on Ligra's data-parallel primitives and adaptive traversal model to process bucketed or frontier-based vertex peeling in parallel.

Dhulipala et al. introduce Julienne~\cite{Julienne2017}, a shared-memory framework extending Ligra with a work-efficient parallel bucketing interface for graph algorithms that rely on bucketed vertex processing, such as k-core decomposition. They design and analyze a parallel bucketing data structure that supports dynamic bucket updates in $O(|V| + |E|)$ expected work and $O(\log n)$ depth per iteration, enabling the first work-efficient parallel k-core algorithm with nontrivial parallelism.
 
A recent work by Liu et al.~\cite{Liu_Theory_and_Practice_2025} presents a new work-efficient parallel k-core decomposition framework that unifies and simplifies prior approaches while achieving strong practical scalability. They prove that a simple frontier-based formulation attains $O(|V| + |E|)$ work without the complex bucketing structures used in earlier systems (such as Julienne~\cite{Julienne2017}). To increase parallelism, the authors introduce a probabilistic sampling scheme that limits contention on high-degree vertices, and vertical granularity control to reduce synchronization overhead in sparse graphs, plus a hierarchical bucketing structure for improved load balance. 

\subsection{Montresor et al. Algorithmic Protocol}
In this subsection we briefly present the protocol that we exploited in our implementation, originally by Montresor et al. \cite{montresor}.

\begin{algorithm}[htb]
    \DontPrintSemicolon
    \caption{Montresor et al. algorithm to compute coreness of node $u$ \cite{montresor}.}
    \label{alg:montresor}
    \textbf{on initialization do} \;
    core $\gets d(u)$\;
    \lForEach{$v \in \text{neighbors}(u)$}{
        est[v] $\gets \infty$
    }
    \textbf{send} $\langle u, core \rangle$ \textbf{to} $\text{neighbors}(u)$\;
    \textbf{on receive} $\langle v, k \rangle$ do \;
    \If{k $<$ est[v]}{
        est[v] $\gets k$\;
        $t \gets \text{computeCoreness}(est, u, core)$\;
        \lIf{t $<$ core}{
            core $\gets t$;
            changed $\gets \text{true}$
        }
    }
    \textbf{repeat} \textit{every round/iteration} \;
    \If{changed}{
        send $\langle u, core \rangle$ to $\text{neighbors}(u)$\;
        changed $\gets \text{false}$\;
    }
\end{algorithm}

Algorithm~\ref{alg:montresor} is a message-exchange, fully decentralized, approach in which the computation is organized in rounds or \emph{iterations}. For each iteration, the nodes communicate estimates of their coreness value to their neighbors, which, in turn, adjust theirs based on what they received.
%
%
There are two main parts into which we can subdivide the algorithm that runs from every node present in the graph: \emph{initialization}, and \emph{reception of a message}, and we analyze them separately.
\paragraph{Initialization} For each node, we assign its coreness estimate to its degree, as this is the sole information available at creation, set its neighbors' estimates to infinity since they are unknown, and then the node communicates its degree to its neighbors.
\paragraph{Message received} The last event that we consider is the reception of a message from a neighbor, which triggers an update in the table of estimates of the recipient and possibly a change in the coreness value. 
 
Algorithm~\ref{alg:montresor} can be easily extended to the case where nodes are partitioned with respect to a set of host machines.
This is a more realistic scenario, where we can exploit the locality of data: in particular, we can assume to have access to the data structures of each node in the same group, and directly manipulate them, instead of having to wait for each node to complete its computation. 
How to partition the nodes into these sets goes beyond the scope of this paper, and it is a hard problem in general. 
For the purpose of this paper, we will assume to work in a \emph{one-host-multiple-node} environment, and draw inspiration for our optimizations with this knowledge in mind.

\section{Parallel Rust Implementation}\label{sec:ParallelK}
In this section, we discuss the details of our implementation of Algorithm~\ref{alg:montresor}.
As pointed out earlier, we developed 3 versions of the algorithm: the first one is a simple, sequential implementation, used both for testing the correctness and to serve as the baseline for time measurements, which we call \emph{SequentialK}.
Secondly, we implemented a parallel version, called \emph{ParallelK}, that exploits the parallelism tools offered by the external library (i.e. \emph{crate}) \emph{Rayon} and a version that uses native Rust threads.
Finally, we developed a highly optimized version of the code, based on the one-host-multiple-nodes paradigm presented in Section~\ref{sec:sota}. 
The latter version will be our main focus during the experimental phase.

\paragraph{SequentialK}
This version also serves as the basis for the next ones, defining the main data structures and methods. In particular, each node of the graphs maintains: a unique identifier (\emph{id}), its current estimate of the coreness value, its neighbors list, a hash map storing the most recent coreness estimates known for each neighbor, and a message queue that holds the updates received from its neighbors waiting to be processed.
Nodes interact exclusively via message exchanges, mimicking the decentralization of the protocol proposed by Montresor et al.~\cite{montresor}.
Each message, for each node, contains a pair \texttt{(id, coreness estimate)}, and is processed separately by the nodes that receive it.
In fact, despite our simulations happening sequentially and centralized, this design enforces the same constraints as in a real distributed environment, where no node has global access to the graph, and updates propagate only through neighbor-to-neighbor interactions. 
From an algorithmic perspective, the functioning of SequentialK follows the iterative message-passing protocol of the distributed algorithm. During each round, nodes compute new coreness estimates based on their neighbors’ current states, enqueue update messages, and then process the incoming queues. This iterative refinement continues until all nodes stabilize, ensuring convergence to the correct k-core decomposition, due to the convergence arguments proved in \cite{montresor}.
This implementation also highlights one of Rust's unique challenges: since the language enforces strict ownership and borrowing rules, creating a graph with circular, mutable references, necessary for bidirectional message passing, requires careful design. Achieving safe yet flexible references was thus a non-trivial step in building SequentialK.
We remark that SequentialK is deliberately naive, but its role is crucial, as it both ensures that the protocol was correctly implemented, and it offers a baseline for the next performance comparisons.

\paragraph{ParallelK}
This version builds a natural parallelization paradigm on top of SequentialK.
ParallelK, in turn, was built with three distinct sub-strategies, specifically using:
\begin{enumerate}
    \item parallel iterators only,
    \item thread pools,
    \item Rust native thread management.
\end{enumerate}
The first two instances were developed using the external library \emph{Rayon}, which offers a good trade-off between simplicity of code writing and performance.
The first variation with respect to the sequential version is the adoption of an auxiliary \emph{MessageMail} data structure, which contains a FIFO queue to handle the message exchange among nodes in the graph. 
Access to the queue is protected by a Mutex, and each node maintains its own instance of MessageMail: this is useful in order to avoid locking the entire queue every time, and instead, one can lock only the queue belonging to the involved neighbors.
Using the Rayon library, this version of ParallelK is implemented via the \texttt{par\_iter\_mut()} method, which parallelizes the main loops on the vertices and on the neighbors vectors so that each vertex can independently send its messages. 
Once all messages of one round have been processed by every node, the code checks, with a parallel logic \emph{or} operation, if there is at least one node that updated its coreness estimate and thus needs to send other messages in the next iteration of the algorithm. 
If no vertex needs to send updates, the algorithm terminates its execution.
 
With the explicit use of thread pools, in the second version of ParallelK, the vertices of $G$ are partitioned into fixed-sized chunks that are then assigned as a task to the thread pool.
Each thread will then loop over its assigned partition of nodes, collecting their messages and dispatching them to the proper neighbors. 
In this version, we adopted one atomic boolean variable that is the result of the logical or among all nodes: if at least one node needs to send more messages, it is set to true directly by it; otherwise, it is left unchanged.
 
The last ParallelK variant makes use of the native thread library provided by Rust.
This implementation required manual chunking or \emph{batching} of the nodes into groups of fixed size, where the size is defined by a parameter that will be evaluated later during the experimental phase.
The parallel scheme adopted consists in creating many worker threads, and synchronizing them on a barrier, as long as there is the need of doing more iterations. 
The code executed by the main thread is summarized in Algorithm~\ref{alg:master_thread}, while the code executed by the worker threads is listed in Algorithm~\ref{alg:worker_threads}.

\begin{algorithm}[htb]
\caption{ParallelK \texttt{Main} thread}
\DontPrintSemicolon
\label{alg:master_thread}
\KwData{\texttt{nodes}: graph nodes, \texttt{T}: number of workers, \texttt{B}: batch/chunk size}

\texttt{Barrier} $\gets$ new synchronization barrier for \texttt{T+1} thread\;
\texttt{go\_on} $\gets$ \texttt{true}\;
\For{\texttt{i} $\gets 0$ \KwTo \texttt{T-1}}{
    \texttt{spawnThread(WorkerThread(index, nodes, B, Barrier))}
}
\While{\texttt{go\_on}}{
    \texttt{go\_on} $\gets$ \texttt{false}\; 
    \tcp{Wait on the barriers for each workers' phase}
    \texttt{Barrier.wait()}\;
    \texttt{Barrier.wait()}\;
    
    \texttt{index.atomic\_store(0)}\; 
    
    \texttt{Barrier.wait()}\;
}
\texttt{joinAllThreads()}\;
\end{algorithm}

\begin{algorithm}[htb]
\caption{ParallelK \texttt{Worker} Thread}
\DontPrintSemicolon
\label{alg:worker_threads}
\KwIn{\texttt{index}: batch index start, \texttt{nodes}, \texttt{B}: batch size, \texttt{Barrier}, \texttt{go\_on}}
\While{\texttt{go\_on}}{
    \texttt{changed} $\gets$ \texttt{false}\;
    \texttt{start} $\gets$ \texttt{index.fetch\_add(B)}\;
    \lIf{\texttt{start} $\geq$ $|$\texttt{nodes}$|$}{
        \textbf{break}
    }
    \texttt{end} $\gets$ $\min$(\texttt{start + B}, $|$\texttt{nodes}$|$)\;
    
    \For{\texttt{u} $\gets$ \texttt{start} \KwTo \texttt{end - 1}}{
        \texttt{u.messages.lock()}\;
        \While{\texttt{u.messages.is\_empty()}}{
            (\texttt{v}, \texttt{k}) $\gets$ \texttt{u.messages.pop()}\;
            \texttt{u.est[v] = k}\;
        }
        \texttt{u.messages.unlock()}\;
        \texttt{u.compute\_coreness()}\;
        \lIf{\texttt{u.changed}}{
            \texttt{changed} $\gets$ \texttt{true}
        }
    }
    
    \texttt{Barrier.wait()}\;
    
    \For{\texttt{u} $\gets$ \texttt{start} \KwTo \texttt{end - 1}}{
        \If{\texttt{u.changed}}{
            \texttt{send\_messages(u)}\;
            \texttt{u.changed} $\gets$ \texttt{false}\;
        }
    }
    \texttt{go\_on.atomic\_or(changed)}\;
    \texttt{Barrier.wait()}\;
}
\end{algorithm}

\subsection{Optimizations}
The parallel versions have gone through a heavy optimization stage, conducted with the Samply \cite{samply_profiler} profiler for Rust, using instruments like the flamegraph and the callgraph. These plots highlight the \emph{hotspots} and bottlenecks of the code, also from the cache point of view, allowing us to carefully tune it to achieve the best performance possible.
We list below the changes introduced in this phase.
\paragraph{Sorted Vectors} We switched the representation of the neighborhood of a node from a HashMap to a sorted vector in order to speed up the computations related to the neighbors.
While this may seem counterintuitive, as hash maps offer $O(1)$ asymptotic complexity for lookup, we actually benefit from vectors being contiguous in memory, and thus in the cache. 
As the degree of real-world nodes in graphs is usually small (see Table~\ref{tab:dataset}), in practice we are able to achieve better performances by doing binary search on a sorted vector, or sometimes even a linear scan, thanks to the cache-friendliness of these operations (on small vectors) \cite{Rucci:2023CAGE}. 
Moreover, we save the time required to compute the hashing function.
Later in the experiments, we will show that this optimization led to up to 30\% faster execution times.
 
\paragraph{Single Rounds} As we previously explained, each iteration of the algorithm is essentially divided into two smaller parts: one that processes the messages received and updates internal structures, and one where each node sends new updates.
This organization of the code requires using two barriers in practice, at the end of each sub-round.
To avoid this, we can join the two phases by sending a new message right after updating one's coreness estimate, hence reducing the synchronization cost, improving locality of cache data, and reducing the overall number of iterations needed to converge to the solution.
 
\paragraph{Selective Message Sending} The protocol presented in Algorithm~\ref{alg:montresor} is blind with respect to the utility of messages sent, i.e., it is not aware if a message will trigger a change in the recipient's estimate. 
A further optimization is possible, assuming each node has local access to the last estimates received by its neighbors, which is checking if $estimate(u) < estimate(v)$
when sending an update from $u$ to $v$: if it is true, then the message is sent as usual; otherwise, that message is discarded, because the coreness value of a node cannot increase during the algorithm execution \cite[Theorem 5]{montresor}.
This optimization allows us to reduce the total number of messages exchanged, and to reduce the number of lock-unlock operations needed, thus reducing also the synchronization overhead.

\section{Improved Algorithm: FastK}\label{sec:FastK}
In the previous section, we explored implementations based on optimizing the basic algorithm, which is based on the ``think like a vertex'' paradigm, where each vertex acts on its own, and possibly on separate machines, which is suitable for distributed infrastructures.
However, this scenario is not realistic in the days of data deluge, and it is instead more realistic to have subsets of vertices spread across a few machines.
Adopting the one-host-multiple-nodes paradigm, we were able to develop a version of the code much more optimized and faster. We present the main points below.
\paragraph{Data Structures} Unlike ParallelK, FastK more effectively leverages the shared memory architecture of the machines, by making the vectors \texttt{est} and \texttt{active} global, shared by all threads and thus visible to all vertices.
These vectors contain, respectively, the current coreness estimate of vertex $i$ in \texttt{est[i]}, and a boolean indicating if vertex $i$ needs to recompute its coreness during the next round, i.e., if it has received a message by at least one of its neighbors.
To reduce the overhead introduced by these new global data structures, we adopted the usage of \emph{raw pointers}: these pointers, which behave much like C++ pointers, allowed us to remove the overhead introduced by synchronized wrappers like Arcs and Mutexes. 
The Rust compiler, however, considers this behavior to be \emph{unsafe}, as it cannot impose the same security checks on raw pointers. We accept these risks, and we carefully orchestrated the code to be as safe as possible, using all the synchronization primitives to ensure no deadlocks happen.
Finally, we note that these improvements also allow for a reduction in auxiliary space occupation, as the vector of estimates is now shared across all vertices, and each vertex does not keep a copy of the estimates for each of its neighbors.
 
\paragraph{Iteration breakdown and synchronization}
Just like ParallelK, FastK keeps each iteration structured in two phases: the first receives messages and updates the local estimates, and the latter sends new messages.
These two phases are outlined in Algorithm~\ref{alg:fastk_worker}, where again we make use of a synchronization barrier in order to wait for all vertices to be consistently updated before they send a new message.

\begin{algorithm}[htb]
\caption{\texttt{FastK}: Worker Thread Logic}
\DontPrintSemicolon
\label{alg:fastk_worker}
\KwIn{Shared \texttt{chunks}, vectors \texttt{est}, \texttt{pending}, \texttt{barrier}, atomic boolean \texttt{changed}}

\While{\texttt{true}}{
    \texttt{local\_updates} $\gets$ empty list\;
    \texttt{local\_activated} $\gets$ empty list\;
    \texttt{local\_changed} $\gets$ \texttt{false}\;

    \texttt{barrier.wait()} \tcp*{Process phase}

    \While{$\exists$ chunk to be processed}{
        \texttt{chunk} $\gets$ get a chunk\;

        \ForEach{\texttt{node} in \texttt{chunk}}{
            \If{\texttt{active[node.id]} = \texttt{false}}{\textbf{continue}}
            \texttt{active[node.id]} $\gets$ \texttt{false}\;
            \If{\texttt{compute\_coreness(node, est)}}{
                \texttt{local\_updates.push((node.id, node.coreness))}\;
                \texttt{local\_changed} $\gets$ \texttt{true}\;

                \ForEach{\texttt{v} in \texttt{node.neighbors}}{
                    \If{\texttt{est[v]} $>$ \texttt{node.coreness}}{
                        \texttt{local\_activated.push(v)}
                    }
                }
            }
        }
    }

    \texttt{changed.fetch\_or(local\_changed)}\;
    \texttt{barrier.wait()} \tcp*{Update phase}

    \ForEach{(\texttt{id}, \texttt{core}) in \texttt{local\_updates}}{
        \texttt{est[id]} $\gets$ \texttt{core}\;
    }
    \ForEach{\texttt{id} in \texttt{local\_activated}}{
        \texttt{active[id]} $\gets$ \texttt{true}\;
    }

    \texttt{barrier.wait()} \tcp*{Iteration end}
    \lIf{\texttt{!changed}}{\textbf{break}}
}
\end{algorithm}%

\paragraph{Further Optimizations}
In addition to the optimizations integrated into ParallelK, FastK exploits two more key optimizations.
Suppose that node $u$ changes its coreness estimate, and let $v$ be a neighbor of $u$.
This change will alter the estimate of $v$ only if either $estimate(u) < estimate(v)$ (as in the previous version), or if the old estimate of $u$ is greater or equal to that of $v$.
In fact, if the latter condition is not true, the old estimate of $u$ was not considered by $v$ during its latest coreness estimate computation, implying that $u$ was not among the $estimate(v)$ neighbors of $v$ with coreness greater or equal to $estimate(v)$, thus a change from $u$ would not be used.
This observation can greatly reduce the total number of messages exchanged, but introduces another synchronization barrier in Algorithm~\ref{alg:fastk_worker}.
 
During the execution of FastK, we observed a recurring pattern during the activation of nodes (i.e., when a node receives a message, it becomes active in the next round). 
In particular, the number of activated nodes decreases rapidly during the first few iterations, then it stabilizes on low values, and this usually leads to slower convergence (see Section~\ref{subsec:convergence}).
We exploit this by dynamically adjusting the parallelization rate, using more threads in the first rounds of the algorithm, then reverting to a sequential-like execution by putting the remaining vertices into a priority queue where the priority is given by the (lowest) coreness estimates.
This helps in avoiding recomputing the coreness of nodes with high coreness too frequently due to low-degree neighbors changing estimates more often.
The condition to check in order to go from parallel to sequential computation has been empirically established during our experiments, and we found that the best tradeoff was given by the proportion of current active nodes with respect to the graph and batch size. 
When this number is less than the batch size (recall that nodes are assigned in batched groups to threads), then we make the switch.

\section{Experimental Results}\label{sec:exp}

\subsection{Dataset, Environment, and Methodology}
We extensively tested our implementations on a machine equipped with two x86\_64 AMD EPYC 7551 processors, with 32 cores each, 128GB of shared RAM. The number of threads used varied as $t = 1, 2, 4, \dots, 128$.
Our Rust code was compiled with the \texttt{--release} flag, which allows for major code optimizations and LTO (Link Time Optimization).
The dataset we used comes from the Stanford Network Analysis Projects (SNAP) Repository and contains real-world data with varying sizes, ranging from thousands to millions of nodes and edges. 
A portion of the dataset used is presented in Table~\ref{tab:dataset}. 
Some of the graphs were originally directed and/or contained self-loops; for the purpose of these experiments, we treated the graph as undirected and removed the self-loops.

\begin{table}[htb]
\caption{Subset of our dataset, sorted by number of edges. $k_{max}$ is the maximum coreness value, $k_{avg}$ the average, and $d_{avg}$ is the average degree.}\label{tab:dataset}
\centering
\resizebox{\columnwidth}{!}{%
\begin{tabular}{|l|r|r|r|r|r|}
\hline
\multicolumn{1}{|c|}{\textbf{Graph}} & 
\multicolumn{1}{c|}{\textbf{$|V|$}} & 
\multicolumn{1}{c|}{\textbf{$|E|$}} & 
\multicolumn{1}{c|}{\textbf{$k_{\text{max}}$}} & 
\multicolumn{1}{c|}{\textbf{$k_{\text{avg}}$}} & 
\multicolumn{1}{c|}{\textbf{$d_{\text{avg}}$}} \\ \hline
roadNet-PA     & 1.088.092 & 1.541.898  & 3   & 1.80  & 2.83  \\ \hline
roadNet-TX     & 1.379.917 & 1.921.660  & 3   & 1.79  & 2.76  \\ \hline
roadNet-CA     & 1.965.206 & 2.766.607  & 3   & 1.81  & 2.81  \\ \hline
web-NotreDame  & 325.729   & 1.497.134  & 155 & 4.32  & 6.69  \\ \hline
web-Stanford   & 281.903   & 2.312.497  & 71  & 7.91  & 14.14 \\ \hline
web-Google     & 875.713   & 5.105.039  & 44  & 5.94  & 9.43  \\ \hline
Wiki-Talk      & 2.394.385 & 5.021.410  & 131 & 1.96  & 3.89  \\ \hline
web-BerkStan   & 685.230   & 7.600.595  & 201 & 11.11 & 19.41 \\ \hline
soc-Pokec      & 1.632.803 & 30.622.564 & 47  & 13.93 & 27.32 \\ \hline
soc-LiveJournal& 4.847.571 & 68.993.773 & 372 & 9.38  & 17.68 \\ \hline
\end{tabular}%
}
\end{table}

We conducted experiments on SequentialK, ParallelK, and FastK, using NetworkX~\cite{NetworkX} as a baseline for correctness verification and timing comparison. The results presented below are the average of 5 executions of the algorithms, which do not take into account the time needed for loading the graph, but include the time for preparing the parallelization-related tasks.
The main metrics gathered consist of running time and convergence speed, measured as the difference, for each node, between its coreness estimate and its final (correct) coreness value during each iteration.

\subsection{Parameter Tuning}
We firstly tested the impact of the different parameters that we introduced, such as the batch size, and the impact of various implementation choices, like vectors/hashmaps and parallel iterators vs thread pools, to decide which version to keep in the subsequent experiments.

\begin{figure}[htb]
    \centering
    \begin{subfigure}[b]{0.5\columnwidth}
        \includegraphics[width=\linewidth]{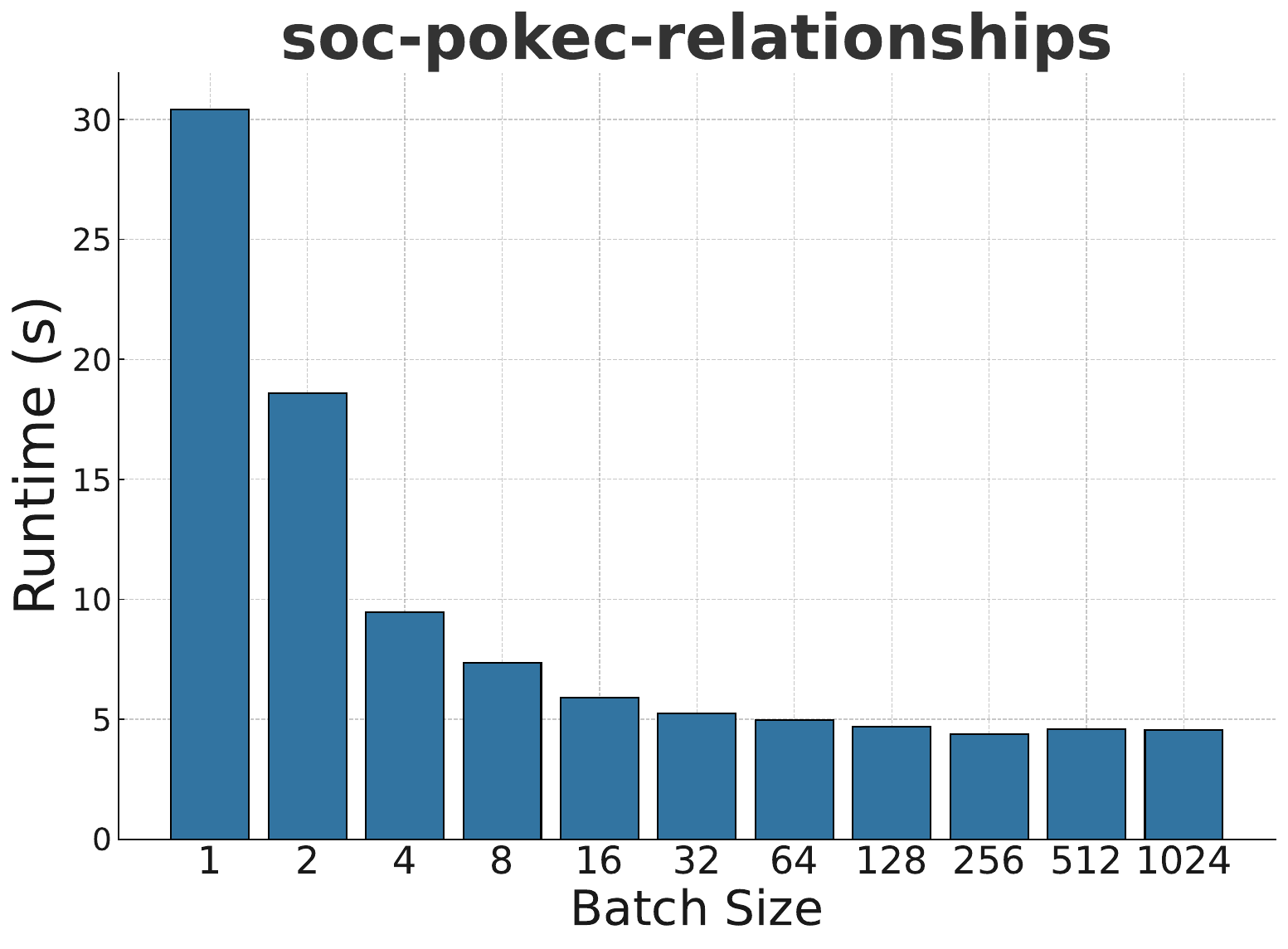}
        \caption{}
        \label{subfig:batch-pokec}
    \end{subfigure}
    \hfill
    \begin{subfigure}[b]{0.45\columnwidth}
        \includegraphics[width=\linewidth]{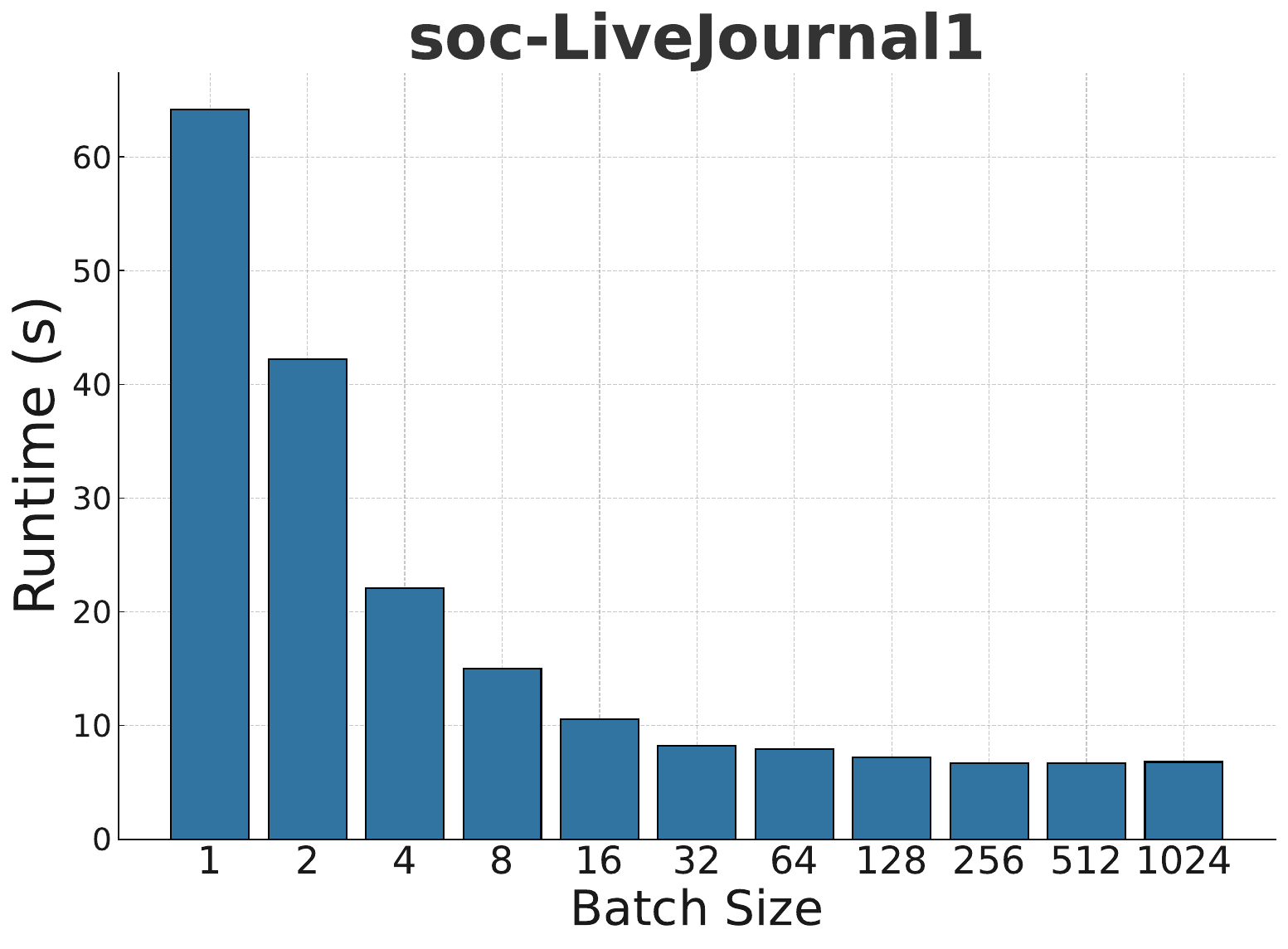}
        \caption{}
        \label{subfig:batch-livejournal}
    \end{subfigure}
    \caption{Runtime with respect to batch size for the ParallelK implementation.}
    \label{fig:batch-social}
\end{figure}

Fig.~\ref{fig:batch-social} shows how the overall running time of ParallelK changes with the increasing batch size. 
The same behavior was observed across the whole dataset; thus, we chose 256 nodes per thread (i.e., batch size) for the subsequent experiments.

\begin{figure}[htb]
    \centering
    \includegraphics[width=\columnwidth]{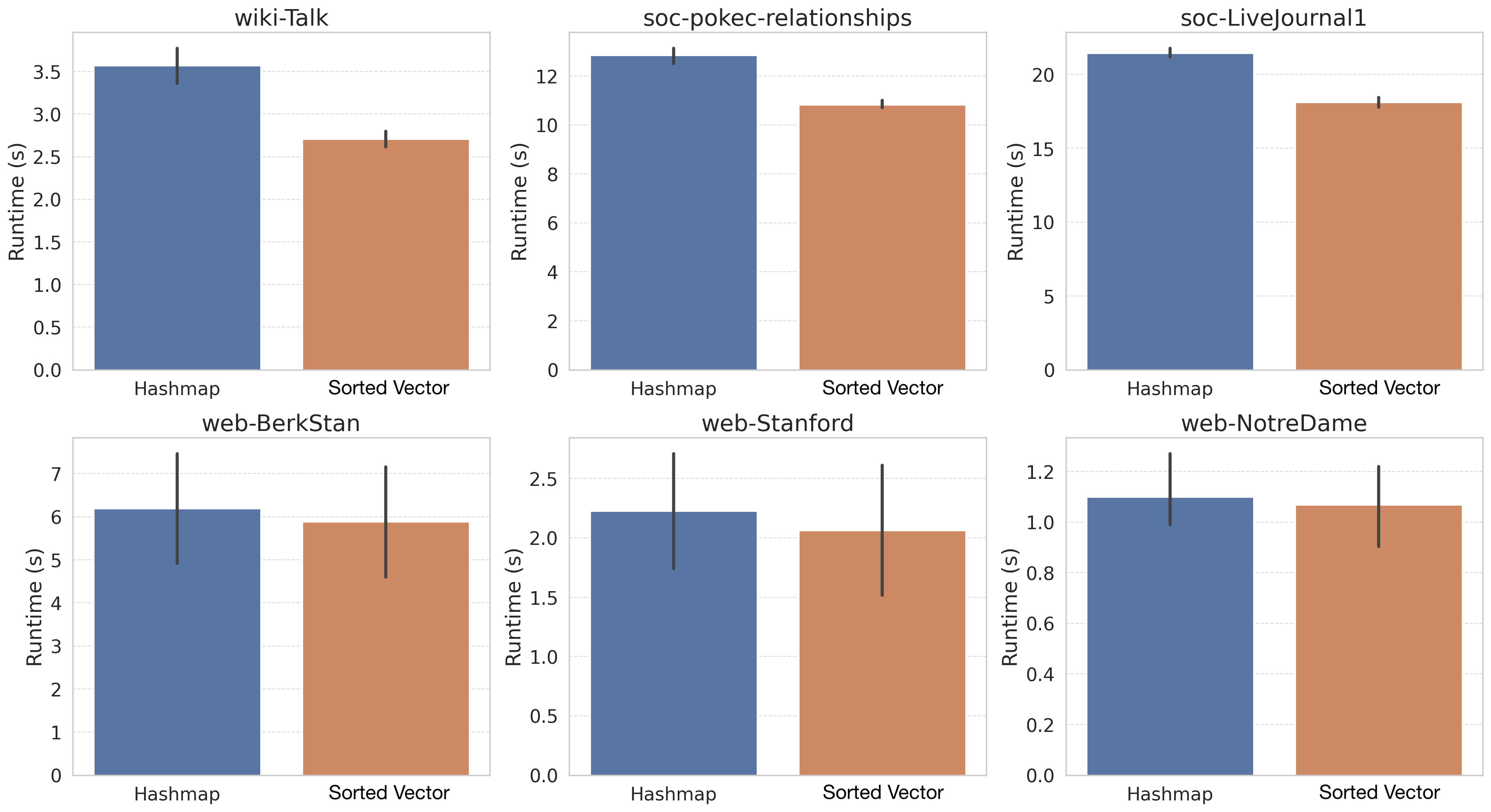}
    \caption{Runtime comparison between implementation with hashmaps and with sorted vectors.}
    \label{fig:array_vs_hashmap}
\end{figure}

Fig.~\ref{fig:array_vs_hashmap} shows the impact on the running time by the neighbors data structure. 
As we motivated earlier in Section~\ref{sec:ParallelK}, we expected an improvement in the running time when using sorted vectors for representing the neighborhood of a vertex because even for large graphs, nodes have low average degree, thus we can benefit from a compact representation like the one of vectors, because it is more cache-friendly.
The plots confirm this hypothesis, allowing us to save some more time over the whole dataset.

\begin{figure}[htb]
    \centering
    \includegraphics[width=\columnwidth]{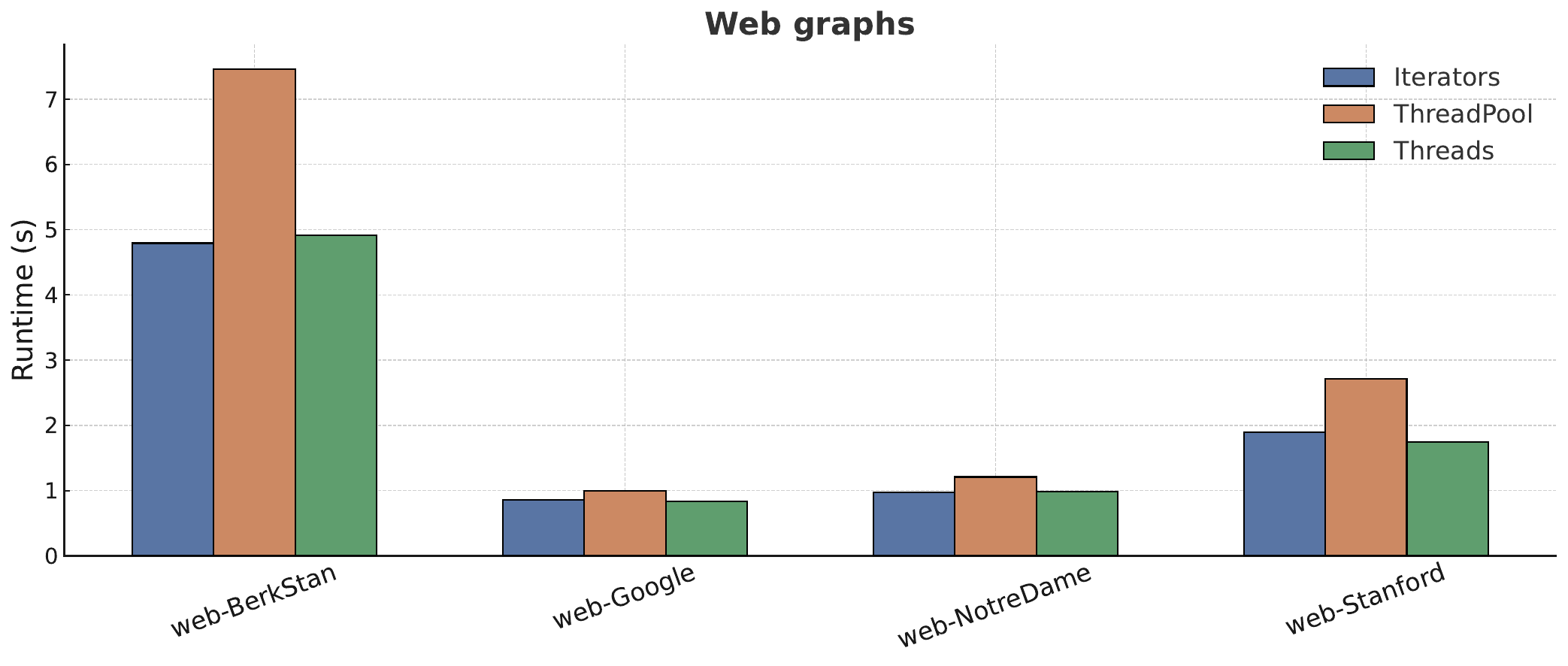}
    \caption{Runtime comparison among different parallelization strategies in ParallelK.}
    \label{fig:thread_runtime}
\end{figure}

Fig.~\ref{fig:thread_runtime} shows the impact of the implementation strategy for the threads between native threads and high-level libraries (i.e., Rayon), with a fixed number of threads (6).
As expected, using an external library introduces management overhead on the computation, thus we chose to use the native thread option for the other experiments, and for developing FastK.

\subsection{Convergence Speed}\label{subsec:convergence}
We now turn to the analysis of convergence speed, i.e., how the coreness values of nodes evolve during the execution of the algorithm.
Since this is neither an approximate algorithm nor a heuristic, the goal of the analysis is not to measure the number of errors made, but rather to understand how close the intermediate results are to the final solution as the algorithm progresses.

\begin{figure}[htb]
    \centering
    \begin{subfigure}[b]{\columnwidth}
        \includegraphics[width=\linewidth]{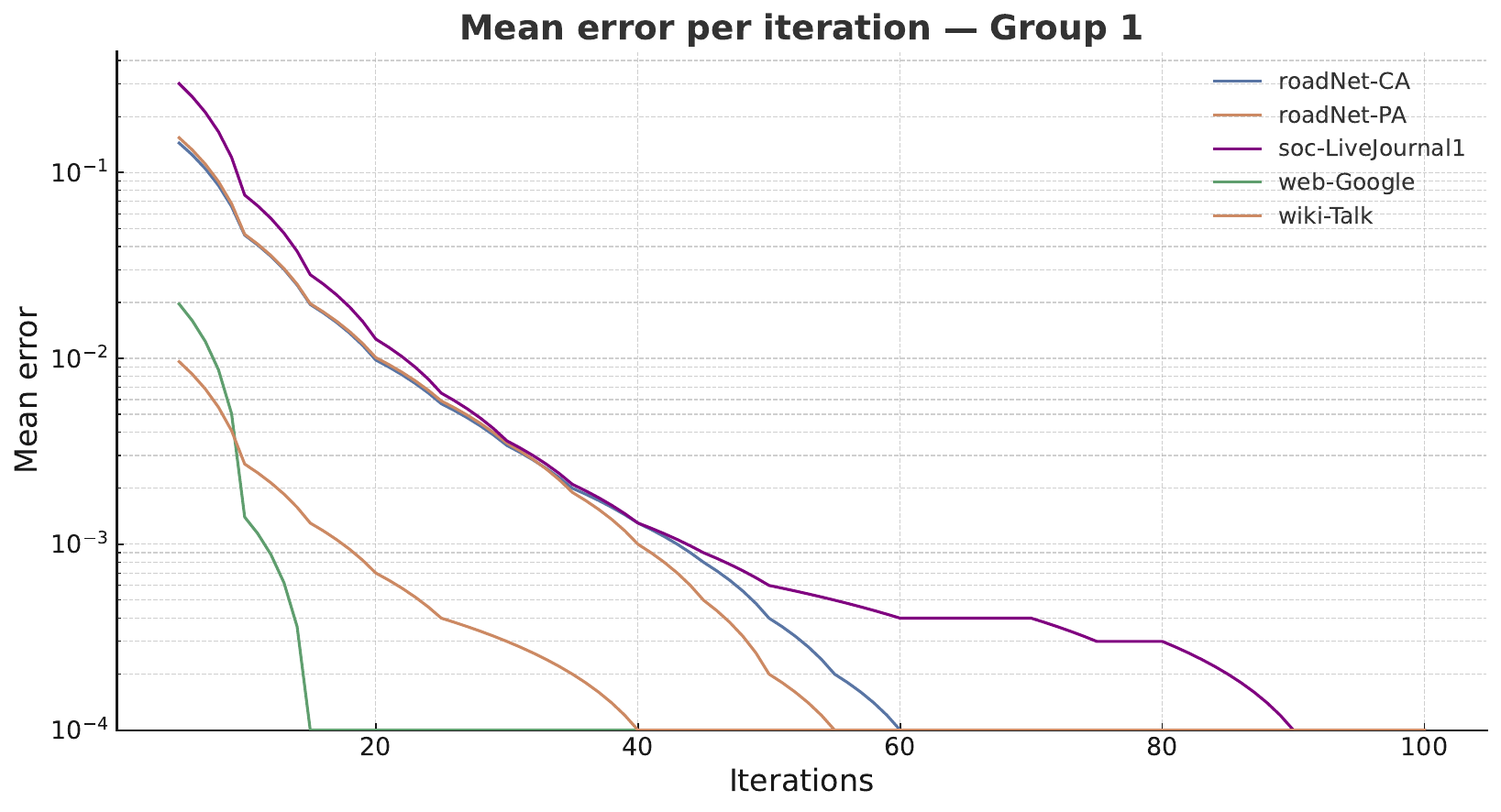}
        \caption{}
    \end{subfigure}
    \\
    \begin{subfigure}[b]{\columnwidth}
        \includegraphics[width=\linewidth]{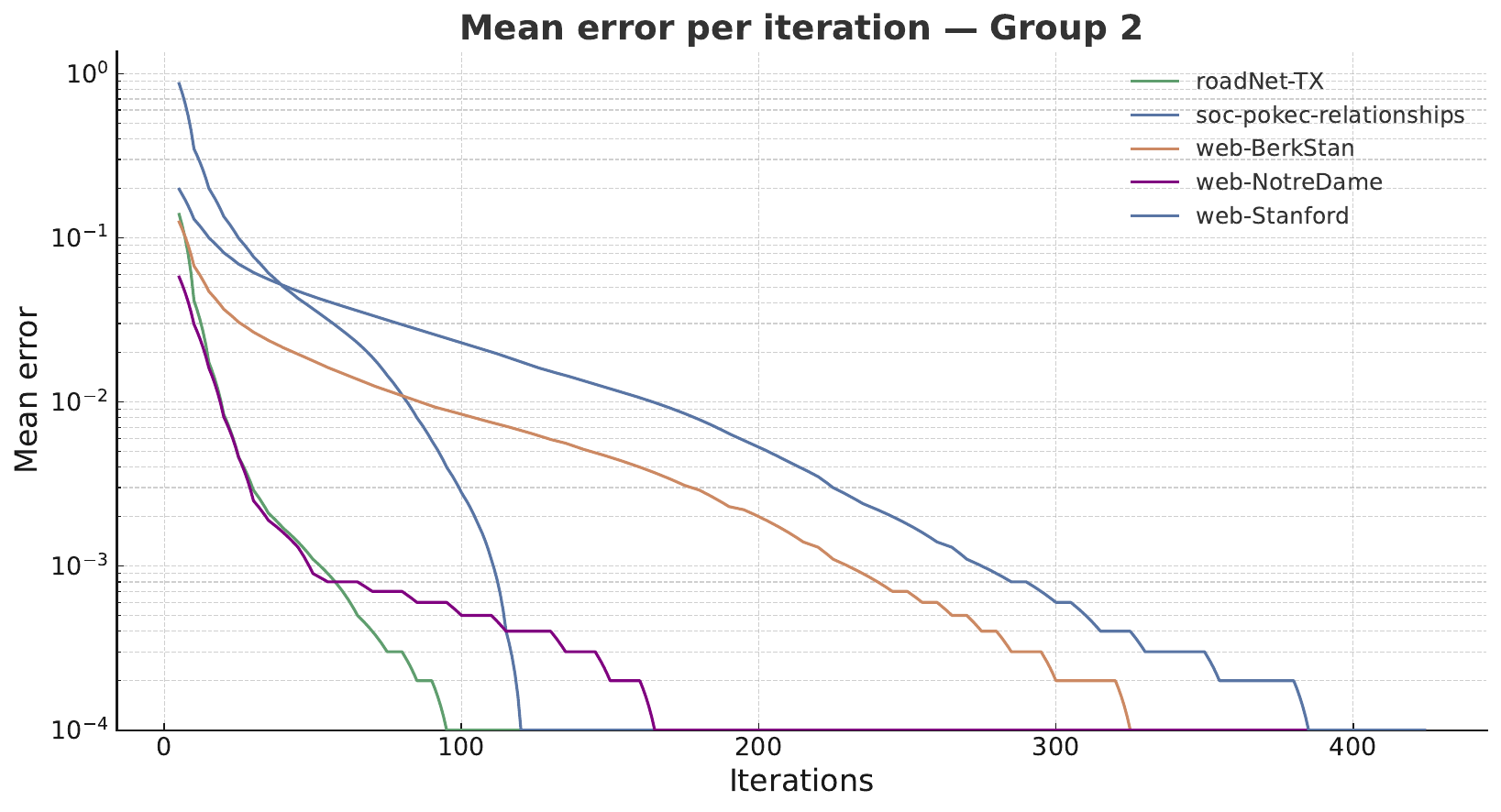}
        \caption{}
    \end{subfigure}
    \caption{Average distance from nodes' true coreness value and the current estimate during the execution of FastK.}
    \label{fig:convergence}
\end{figure}

\begin{figure}[htb]
    \centering
    \includegraphics[width=\columnwidth]{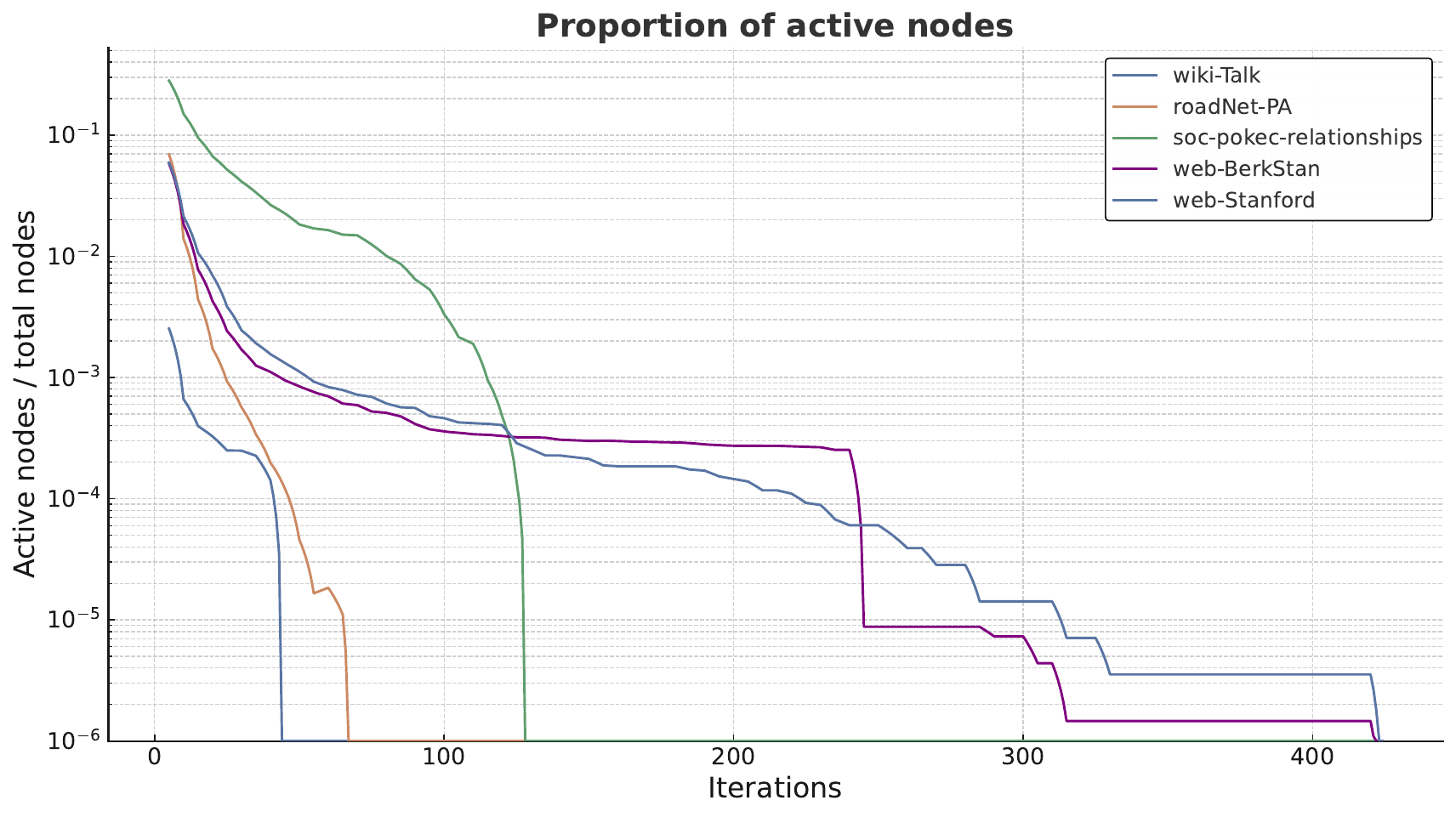}
    \caption{Percentage of activated nodes during the execution of FastK.}
    \label{fig:activated_nodes}
\end{figure}

Fig.~\ref{fig:convergence} shows the convergence speed of FastK, and it highlights how the difference between the true value and the current estimate decreases rapidly during the execution.
Observe, in particular, how the error stays stable towards the end of the execution: this means that the last iterations are somewhat ``fine-tuning'' the result, and most of the vertices deviate $\pm 1$ from their true coreness.
The number of nodes still not completed in this phase is usually small (ref. Fig.~\ref{fig:activated_nodes}), hence the swapping to a sequential execution towards the end (as explained in Section~\ref{sec:FastK}). 

\subsection{Running Time}
\begin{table}[htb]
\caption{Average running times (seconds) for each graph and algorithm. Parallel algorithms use 16 threads.}
    \centering
    \begin{tabular}{|l|r|r|r|r|}
    \hline
    \textbf{Graph} &  \textbf{FastK} &  \textbf{NetworkX} &  \textbf{ParallelK} &  \textbf{SequentialK} \\
    \hline
    roadNet-CA              &   \textbf{0.35} &     40.48 &       0.82 &         2.45 \\
    \hline
    roadNet-PA              &   \textbf{0.22} &      8.85 &       0.49 &         1.53 \\
    \hline
    roadNet-TX              &   \textbf{0.42} &     10.57 &       1.09 &         2.49 \\
    \hline
    soc-LiveJournal1        &   \textbf{3.87} &    480.92 &       7.07 &        78.11 \\
    \hline
    soc-pokec               &   \textbf{3.13} &    161.92 &       4.54 &        53.78 \\
    \hline
    web-BerkStan            &   \textbf{0.59} &    686.63 &       2.22 &         5.19 \\
    \hline
    web-Google              &   \textbf{0.23} &     38.40 &       0.34 &         3.13 \\
    \hline
    web-NotreDame           &   \textbf{0.09} &      6.55 &       0.16 &         0.62 \\
    \hline
    web-Stanford            &   \textbf{0.21} &     93.16 &       0.57 &         2.42 \\
    \hline
    wiki-Talk               &   \textbf{0.61} &    115.28 &       0.82 &         8.76 \\
    \hline
    \end{tabular}
    
    \label{tab:runtime}
\end{table}

\begin{figure}[htb]
    \centering
    \begin{subfigure}[b]{\columnwidth}
        \includegraphics[width=\linewidth]{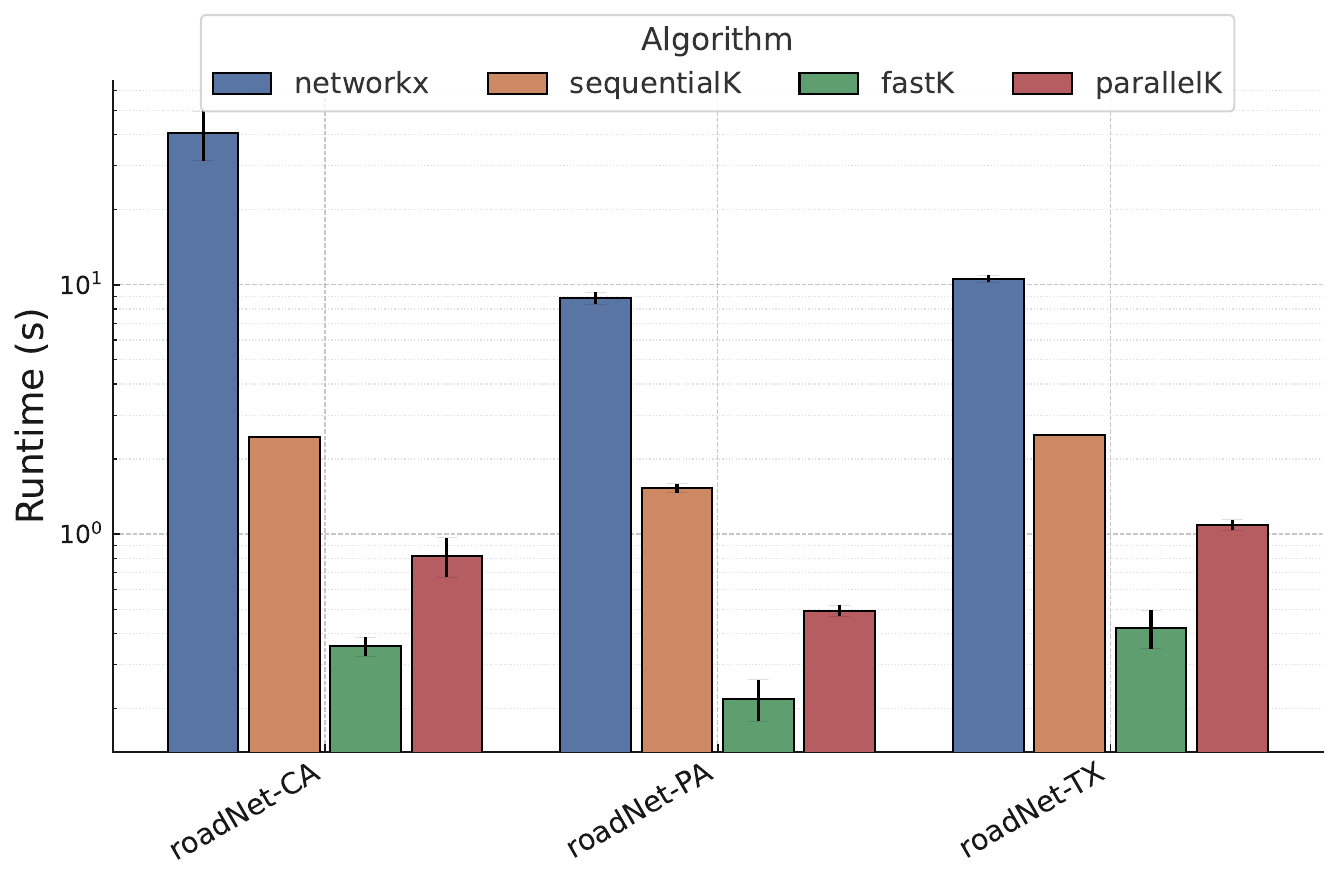}
        \caption{}
    \end{subfigure}
    \\
    \begin{subfigure}[b]{\columnwidth}
        \includegraphics[width=\linewidth]{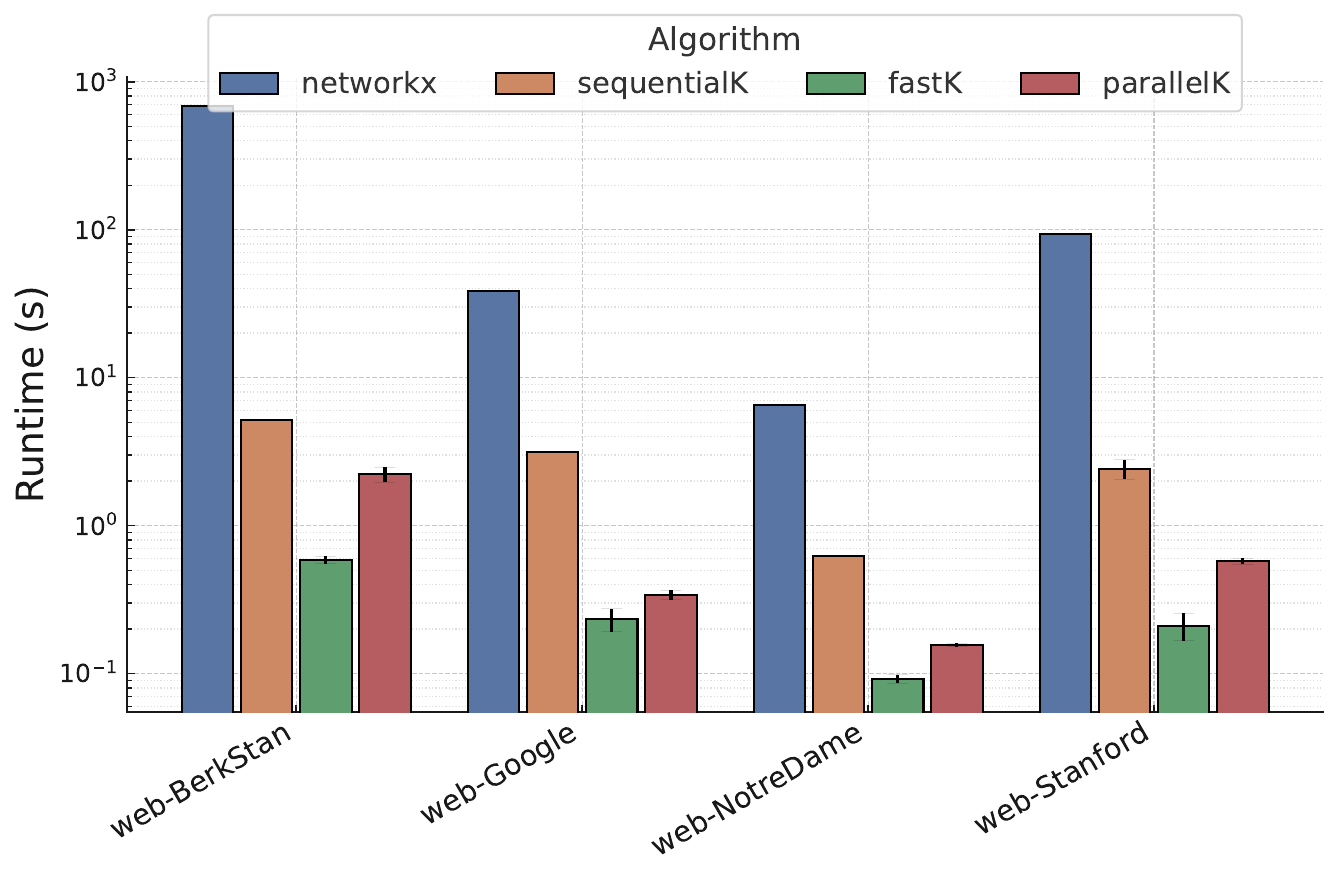}
        \caption{}
    \end{subfigure}
    \caption{Running time comparison for all implementations and NetworkX as a baseline (number of threads: 16).}
    \label{fig:running_time}
\end{figure}

Finally, we compare the overall performance of the algorithms against each other to verify our claims on the speedup introduced by the multiple optimizations that we discussed.
In what follows, the number of threads has been fixed to 16.
We present the results in Table~\ref{tab:runtime} and Fig.~\ref{fig:running_time}.
The results obtained show that FastK consistently outperforms all other variants, and it is many orders of magnitude faster than the (sequential) implementation of NetworkX, even though FastK includes a sequential segment within it.
ParallelK also achieves very good performances, in the same order of magnitude as FastK, although the latter is clearly faster. This validates our work and empirically shows the soundness of our various optimization strategies.

\begin{figure}[htb]
    \centering
    \begin{subfigure}[b]{\columnwidth}
        \includegraphics[width=\linewidth]{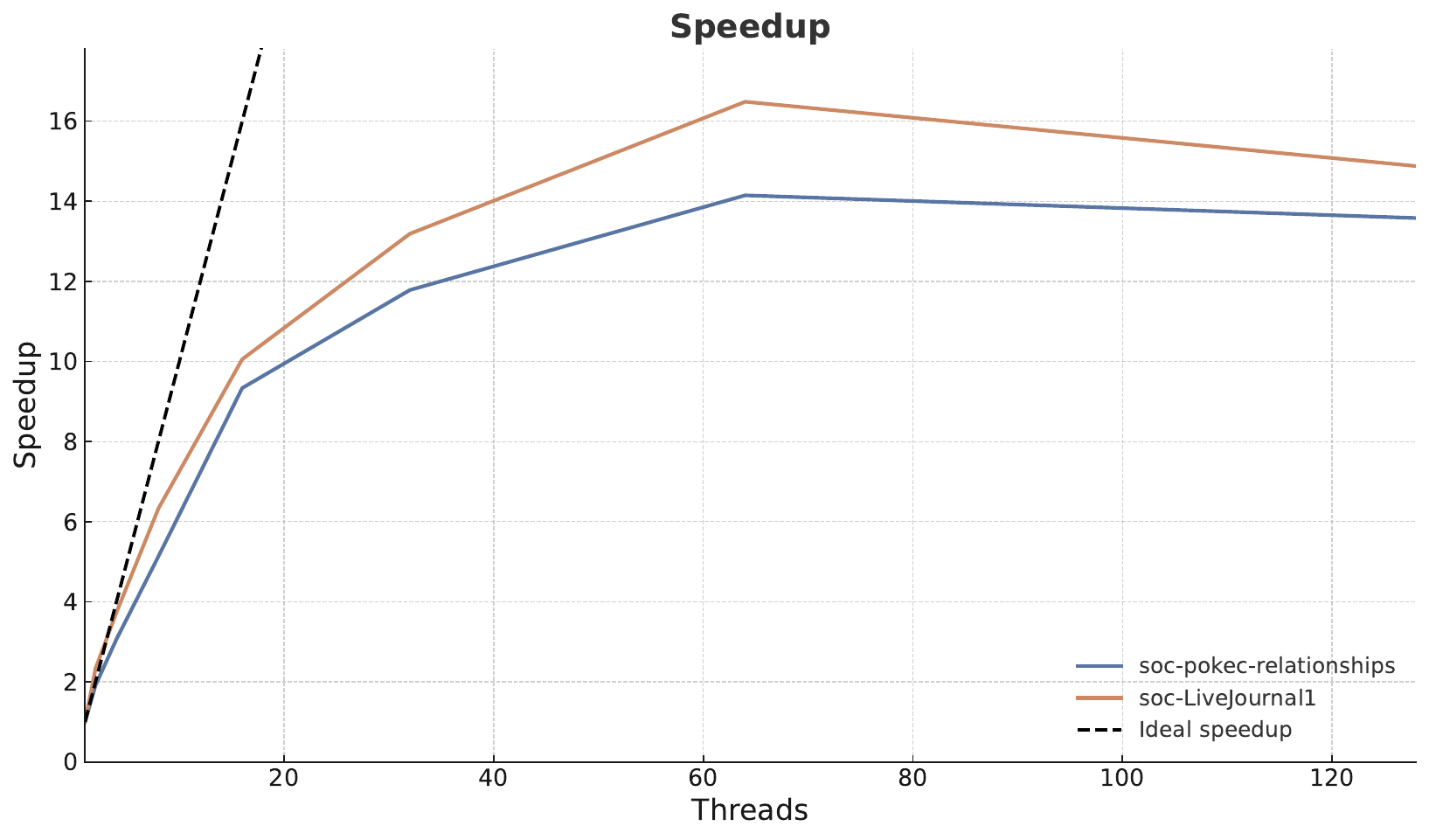}
        \caption{}
    \end{subfigure}
    \\
    \begin{subfigure}[b]{\columnwidth}
        \includegraphics[width=\linewidth]{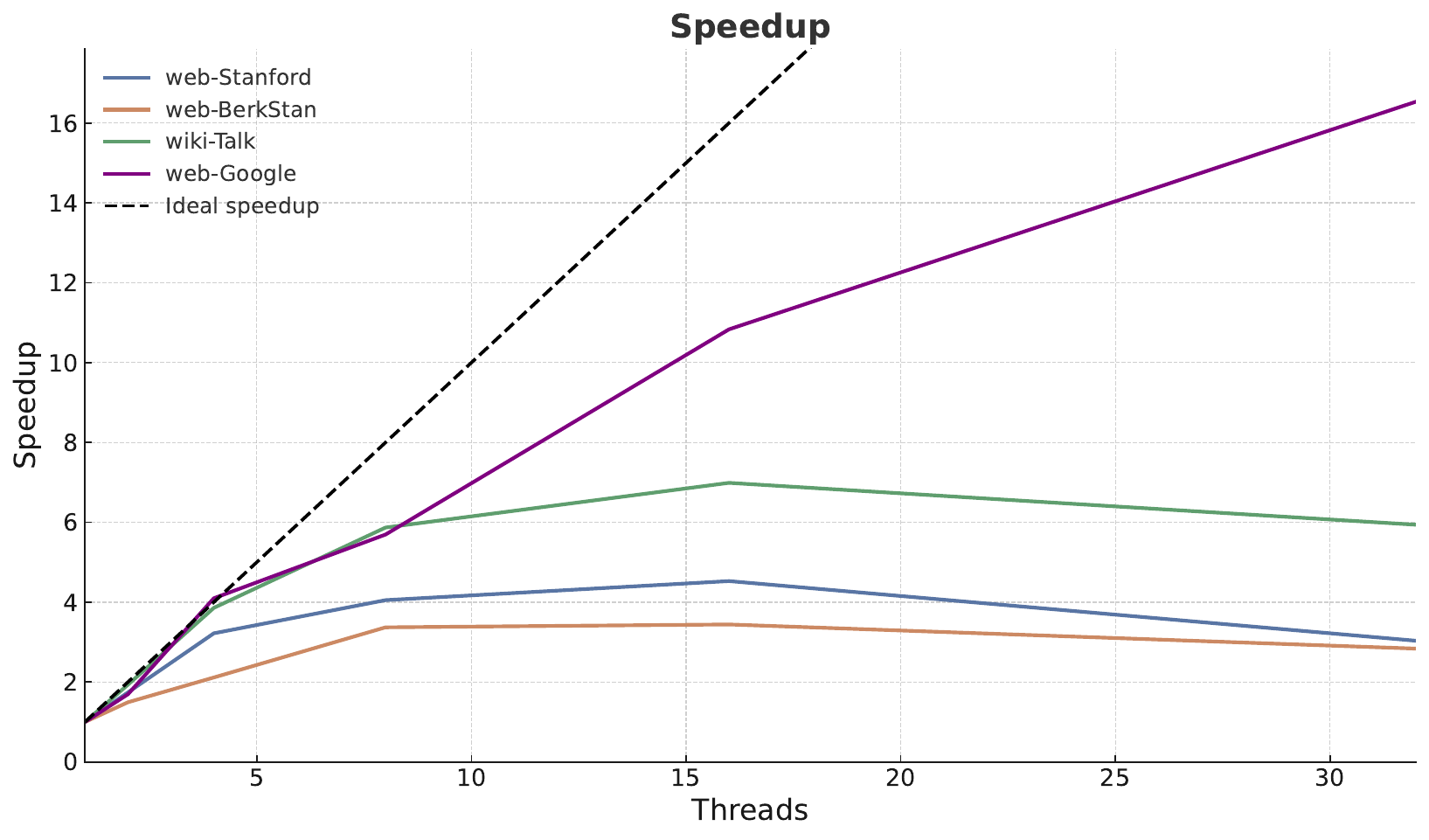}
        \caption{}
    \end{subfigure}
    \caption{Speedup of FastK with respect to the number of available threads.}
    \label{fig:speedup}
\end{figure}

To further highlight the benefits of FastK, we also provide plots for the speedup, obtained as the ratio between the sequential runtime (i.e. with 1 thread) and the runtime with $t$ threads, in Fig.~\ref{fig:speedup}).
As it is always the case with speedup, increasing the number of threads may not always provide a boost in performance, because the overhead introduced by the thread management will be larger as the number of threads grows.
Nonetheless, we can clearly see how the parallelism affects the performance of FastK for the better.

\section{Conclusions}\label{sec:conc}
We presented a Rust-based parallel library for k-core decomposition that adapts a decentralized message-passing protocol to shared-memory multicore systems. Starting from a correctness-oriented baseline (\texttt{SequentialK}), we explored alternative concurrency models (\texttt{ParallelK}) and introduced a cache- and synchronization-aware design (\texttt{FastK}) that leverages global shared state, selective message propagation, and activation strategies. Across large, real-world graphs, \texttt{FastK} achieves up to $11\times$ speedup on 16 threads and runs up to two orders of magnitude faster than \texttt{NetworkX}~\cite{NetworkX}, showing that memory safety and low-level control in Rust can coexist to deliver practical performance.
The results we achieved suggest that the distributed protocol of Montresor \emph{et al.} can be optimized effectively for shared memory when synchronization points are minimized and data structures are cache-friendly. It also gave experimental evidence that message selectivity and activation scheduling reduce contention and unnecessary work, which is critical once only a small fraction of vertices remains active. Moreover, native-thread implementations provided the best trade-off in our setting, but the gain depends on batching and barrier placement.
 
Our evaluation focuses on in-memory, shared-memory execution on a single NUMA machine and excludes graph-loading time from the reported runtimes. Performance may vary with different memory hierarchies, partitioning, or degrees of skewness. Moreover, we studied static graphs; dynamic updates were not evaluated.
 
As future work, we plan to (a) extend the library toward truly distributed deployments that preserve the optimized shared-memory kernels while enabling inter-node communication; (b) investigate dynamic (incremental/decremental) core maintenance; (c) study NUMA-aware partitioning and external-memory variants for graphs exceeding RAM; and (d) explore the exploitation of heterogeneous acceleration (e.g., GPU-assisted filtering).

\bibliographystyle{IEEEtran}
\bibliography{bibliography}

@inproceedings{montresor, author = {Montresor, Alberto and De Pellegrini, Francesco and Miorandi, Daniele}, title = {Distributed k-core decomposition}, year = {2011}, isbn = {9781450307192}, publisher = {Association for Computing Machinery}, address = {New York, NY, USA}, url = {https://doi.org/10.1145/1993806.1993836}, doi = {10.1145/1993806.1993836}, booktitle = {Proceedings of the 30th Annual ACM SIGACT-SIGOPS Symposium on Principles of Distributed Computing}, pages = {207–208}, numpages = {2}, location = {San Jose, California, USA}, series = {PODC '11} }

@article{SEIDMAN1983269,
title = {Network structure and minimum degree},
journal = {Soc. Netw.},
volume = {5},
number = {3},
pages = {269-287},
year = {1983},
issn = {0378-8733},
author = {Stephen B. Seidman}
}

@misc{samply_profiler,
  author       = {Gregor Wagner},
  title        = {Samply: A fast, native sampling profiler for Rust},
  howpublished = {\url{https://github.com/samply}},
  year         = {2022},
  note         = {Accessed: 2025-05-15}
}

@InProceedings{Rucci:2023CAGE,
author="Conte, Alessio
and Grossi, Roberto
and Rucci, Davide",
editor="Nardini, Franco Maria
and Pisanti, Nadia
and Venturini, Rossano",
title="CAGE: Cache-Aware Graphlet Enumeration",
booktitle="String Processing and Information Retrieval",
year="2023",
publisher="Springer Nature Switzerland",
address="Cham",
pages="129--142",
isbn="978-3-031-43980-3"
}

@InProceedings{NetworkX,
  author =       {Aric A. Hagberg and Daniel A. Schult and Pieter J. Swart},
  title =        {Exploring Network Structure, Dynamics, and Function using NetworkX},
  booktitle =   {Proceedings of the 7th Python in Science Conference},
  pages =     {11 - 15},
  address = {Pasadena, CA USA},
  year =      {2008},
  editor =    {Ga\"el Varoquaux and Travis Vaught and Jarrod Millman},
}

@misc{batagelj2003omalgorithmcoresdecomposition,
      title={An O(m) Algorithm for Cores Decomposition of Networks}, 
      author={V. Batagelj and M. Zaversnik},
      year={2003},
      eprint={cs/0310049},
      archivePrefix={arXiv},
      primaryClass={cs.DS},
      url={https://arxiv.org/abs/cs/0310049}, 
}

@INPROCEEDINGS{Park_2014,
  author={Dasari, Naga Shailaja and Desh, Ranjan and Zubair, M.},
  booktitle={2014 IEEE International Conference on Big Data (Big Data)}, 
  title={ParK: An efficient algorithm for k-core decomposition on multicore processors}, 
  year={2014},
  volume={},
  number={},
  pages={9-16},
  doi={10.1109/BigData.2014.7004366}}

@INPROCEEDINGS{PKC,
  author={Kabir, Humayun and Madduri, Kamesh},
  booktitle={2017 IEEE International Parallel and Distributed Processing Symposium Workshops (IPDPSW)}, 
  title={Parallel k-Core Decomposition on Multicore Platforms}, 
  year={2017},
  volume={},
  number={},
  pages={1482-1491},
  doi={10.1109/IPDPSW.2017.151}}

@article{Liu_Theory_and_Practice_2025, author = {Liu, Youzhe and Dong, Xiaojun and Gu, Yan and Sun, Yihan}, title = {Parallel k-Core Decomposition: Theory and Practice}, year = {2025}, issue_date = {June 2025}, publisher = {Association for Computing Machinery}, address = {New York, NY, USA}, volume = {3}, number = {3}, url = {https://doi.org/10.1145/3725332}, doi = {10.1145/3725332}, journal = {Proc. ACM Manag. Data}, month = jun, articleno = {195}, numpages = {27} }

@inproceedings{Julienne2017, author = {Dhulipala, Laxman and Blelloch, Guy and Shun, Julian}, title = {Julienne: A Framework for Parallel Graph Algorithms using Work-efficient Bucketing}, year = {2017}, isbn = {9781450345934}, publisher = {Association for Computing Machinery}, address = {New York, NY, USA}, url = {https://doi.org/10.1145/3087556.3087580}, doi = {10.1145/3087556.3087580}, booktitle = {Proceedings of the 29th ACM Symposium on Parallelism in Algorithms and Architectures}, pages = {293–304}, numpages = {12}, location = {Washington, DC, USA}, series = {SPAA '17} }

@article{Ligra2013, author = {Shun, Julian and Blelloch, Guy E.}, title = {Ligra: a lightweight graph processing framework for shared memory}, year = {2013}, issue_date = {August 2013}, publisher = {Association for Computing Machinery}, address = {New York, NY, USA}, volume = {48}, number = {8}, issn = {0362-1340}, url = {https://doi.org/10.1145/2517327.2442530}, doi = {10.1145/2517327.2442530}, journal = {SIGPLAN Not.}, month = feb, pages = {135–146}, numpages = {12} }

@Article{k_core_survey2020,
author="Malliaros, Fragkiskos D.
and Giatsidis, Christos
and Papadopoulos, Apostolos N.
and Vazirgiannis, Michalis",
title="The core decomposition of networks: theory, algorithms and applications",
journal="The VLDB Journal",
year="2020",
month="Jan",
day="01",
volume="29",
number="1",
pages="61--92",
issn="0949-877X",
doi="10.1007/s00778-019-00587-4",
url="https://doi.org/10.1007/s00778-019-00587-4"
}

@INPROCEEDINGS{faloutsos_corescope,
  author={Shin, Kijung and Eliassi-Rad, Tina and Faloutsos, Christos},
  booktitle={2016 IEEE 16th International Conference on Data Mining (ICDM)}, 
  title={CoreScope: Graph Mining Using k-Core Analysis — Patterns, Anomalies and Algorithms}, 
  year={2016},
  volume={},
  number={},
  pages={469-478},
  doi={10.1109/ICDM.2016.0058}}

@Article{Malvestio2020,
author="Malvestio, Irene
and Cardillo, Alessio
and Masuda, Naoki",
title="Interplay between {\$}{\$}k{\$}{\$}-core and community structure in complex networks",
journal="Scientific Reports",
year="2020",
month="Sep",
day="07",
volume="10",
number="1",
pages="14702",
issn="2045-2322",
doi="10.1038/s41598-020-71426-8",
url="https://doi.org/10.1038/s41598-020-71426-8"
}

@article{santofortunato,
title = {Community detection in graphs},
journal = {Physics Reports},
volume = {486},
number = {3},
pages = {75-174},
year = {2010},
issn = {0370-1573},
doi = {https://doi.org/10.1016/j.physrep.2009.11.002},
url = {https://www.sciencedirect.com/science/article/pii/S0370157309002841},
author = {Santo Fortunato}
}

@article{BAE2014549,
title = {Identifying and ranking influential spreaders in complex networks by neighborhood coreness},
journal = {Physica A: Statistical Mechanics and its Applications},
volume = {395},
pages = {549-559},
year = {2014},
issn = {0378-4371},
doi = {https://doi.org/10.1016/j.physa.2013.10.047},
url = {https://www.sciencedirect.com/science/article/pii/S0378437113010406},
author = {Joonhyun Bae and Sangwook Kim}
}

@INPROCEEDINGS{wu2015core,
  author={Wu, Huanhuan and Cheng, James and Lu, Yi and Ke, Yiping and Huang, Yuzhen and Yan, Da and Wu, Hejun},
  booktitle={2015 IEEE International Conference on Big Data (Big Data)}, 
  title={Core decomposition in large temporal graphs}, 
  year={2015},
  volume={},
  number={},
  pages={649-658},
  doi={10.1109/BigData.2015.7363809}}

@article{kumar2020identifying,
title = {Identifying influential nodes in Social Networks: Neighborhood Coreness based voting approach},
journal = {Physica A: Statistical Mechanics and its Applications},
volume = {553},
pages = {124215},
year = {2020},
issn = {0378-4371},
doi = {https://doi.org/10.1016/j.physa.2020.124215},
url = {https://www.sciencedirect.com/science/article/pii/S0378437120300479},
author = {Sanjay Kumar and B.S. Panda}
}

@inproceedings{Rucci:2024SAC, author = {Conte, Alessio and Rucci, Davide}, title = {Are k-cores meaningful for temporal graph analysis?}, year = {2024}, isbn = {9798400702433}, publisher = {Association for Computing Machinery}, address = {New York, NY, USA}, url = {https://doi.org/10.1145/3605098.3635959}, doi = {10.1145/3605098.3635959}, booktitle = {Proceedings of the 39th ACM/SIGAPP Symposium on Applied Computing}, pages = {1453–1460}, numpages = {8}, location = {Avila, Spain}, series = {SAC '24} }

@article{agouti2022graph,
  title={Graph-based modeling using association rule mining to detect influential users in social networks},
  author={Agouti, Tarik},
  journal={Expert Systems with Applications},
  volume={202},
  pages={117436},
  year={2022},
  publisher={Elsevier}
}

@inproceedings{lin2020guardian,
  title={Guardian: Evaluating trust in online social networks with graph convolutional networks},
  author={Lin, Wanyu and Gao, Zhaolin and Li, Baochun},
  booktitle={IEEE INFOCOM 2020-IEEE Conference on Computer Communications},
  pages={914--923},
  year={2020},
  organization={IEEE}
}

@article{guidi2019towards,
  title={Towards the dynamic community discovery in decentralized online social networks},
  author={Guidi, Barbara and Michienzi, Andrea and Rossetti, Giulio},
  journal={Journal of Grid Computing},
  volume={17},
  number={1},
  pages={23--44},
  year={2019},
  publisher={Springer}
}

@inproceedings{chen2014towards,
  title={Towards scalable graph computation on mobile devices},
  author={Chen, Yiqi and Lin, Zhiyuan and Pienta, Robert and Kahng, Minsuk and Chau, Duen Horng},
  booktitle={2014 IEEE International Conference on Big Data (Big Data)},
  pages={29--35},
  year={2014},
  organization={IEEE}
}

@article{wang2018d2d,
  title={D2D big data: Content deliveries over wireless device-to-device sharing in large-scale mobile networks},
  author={Wang, Xiaofei and Zhang, Yuhua and Leung, Victor CM and Guizani, Nadra and Jiang, Tianpeng},
  journal={IEEE Wireless Communications},
  volume={25},
  number={1},
  pages={32--38},
  year={2018},
  publisher={IEEE}
}

@article{santos2021temporal,
  title={Temporal complex networks modeling applied to vehicular ad-hoc networks},
  author={Santos, Fillipe and Aquino, Andre LL and Madeira, Edmundo RM and Cabral, Raquel S},
  journal={Journal of Network and Computer Applications},
  volume={192},
  pages={103168},
  year={2021},
  publisher={Elsevier}
}

@article{zolghadri2024resource,
  title={Resource allocation in fog--cloud environments: State of the art},
  author={Zolghadri, Mohammad and Asghari, Parvaneh and Dashti, Seyed Ebrahim and Hedayati, Alireza},
  journal={Journal of Network and Computer Applications},
  volume={227},
  pages={103891},
  year={2024},
  publisher={Elsevier}
}

@article{wu2021internet,
  title={Internet of things as complex networks},
  author={Wu, Xing and Wang, Jianjia and Li, Peng and Luo, Xiliang and Yang, Yang},
  journal={IEEE Network},
  volume={35},
  number={3},
  pages={238--245},
  year={2021},
  publisher={IEEE}
}

@article{li2024exploiting,
  title={Exploiting complex network-based clustering for personalization-enhanced hierarchical federated edge learning},
  author={Li, Zijian and Chen, Zihan and Wei, Xiaohui and Gao, Shang and Yue, Hengshan and Xu, Zhewen and Quek, Tony QS},
  journal={IEEE Transactions on Mobile Computing},
  year={2024},
  publisher={IEEE}
}

@article{akhtarshenas2024federated,
  title={Federated learning: A cutting-edge survey of the latest advancements and applications},
  author={Akhtarshenas, Azim and Vahedifar, Mohammad Ali and Ayoobi, Navid and Maham, Behrouz and Alizadeh, Tohid and Ebrahimi, Sina and L{\'o}pez-P{\'e}rez, David},
  journal={Computer Communications},
  volume={228},
  pages={107964},
  year={2024},
  publisher={Elsevier}
}

@article{hong2019resource,
  title={Resource management in fog/edge computing: a survey on architectures, infrastructure, and algorithms},
  author={Hong, Cheol-Ho and Varghese, Blesson},
  journal={ACM computing surveys (csur)},
  volume={52},
  number={5},
  pages={1--37},
  year={2019},
  publisher={ACM New York, NY, USA}
}

@article{sakr2021future,
  title={The future is big graphs: a community view on graph processing systems},
  author={Sakr, Sherif and Bonifati, Angela and Voigt, Hannes and Iosup, Alexandru and Ammar, Khaled and Angles, Renzo and Aref, Walid and Arenas, Marcelo and Besta, Maciej and Boncz, Peter A and others},
  journal={Communications of the ACM},
  volume={64},
  number={9},
  pages={62--71},
  year={2021},
  publisher={ACM New York, NY, USA}
}

@inproceedings{gennaro2007mroute,
  title={MRoute: A peer-to-peer routing index for similarity search in metric spaces},
  author={Gennaro, Claudio and Mordacchini, Matteo and Orlando, Salvatore and Rabitti, Fausto},
  booktitle={Proceedings of the 5th International Workshop on Databases, Information Systems and Peer-to-Peer Computing (DBISP2P 2007)},
  pages={1--12},
  year={2007}
}

@article{mordacchini2017social,
  title={A social cognitive heuristic for adaptive data dissemination in mobile Opportunistic Networks},
  author={Mordacchini, Matteo and Passarella, Andrea and Conti, Marco},
  journal={Pervasive and Mobile Computing},
  year={2017},
  publisher={Elsevier}
}

@inproceedings{ferrucci2021latency,
  title={Latency Preserving Self-optimizing Placement at the Edge},
  author={Ferrucci, Luca and Mordacchini, Matteo and Coppola, Massimo and Carlini, Emanuele and Kavalionak, Hanna and Dazzi, Patrizio},
  booktitle={Proceedings of the 1st Workshop on Flexible Resource and Application Management on the Edge},
  pages={3--8},
  year={2021},
  organization={ACM}
}

@inproceedings{kavalionak2021impact,
  title={Impact of network topology on the convergence of decentralized federated learning systems},
  author={Kavalionak, Hanna and Carlini, Emanuele and Dazzi, Patrizio and Ferrucci, Luca and Mordacchini, Matteo and Coppola, Massimo},
  booktitle={2021 IEEE Symposium on Computers and Communications (ISCC)},
  pages={1--6},
  year={2021},
  organization={IEEE}
}

@article{ferrucci2024decentralized,
  title={Decentralized replica management in latency-bound edge environments for resource usage minimization},
  author={Ferrucci, Luca and Mordacchini, Matteo and Dazzi, Patrizio},
  journal={IEEE Access},
  volume={12},
  pages={19229--19249},
  year={2024},
  publisher={IEEE}
}

@inproceedings{mordacchini2025decentralized,
  title={A decentralized strategy for unweighted minimum vertex cover},
  author={Mordacchini, Matteo and Carlini, Emanuele and Dazzi, Patrizio},
  booktitle={Proceedings of the 40th ACM/SIGAPP Symposium on Applied Computing},
  pages={325--328},
  year={2025}
}

@inproceedings{mordacchini2025ICDCS,
  title={A Fast-Converging Decentralized Approach to the Weighted Minimum Vertex Cover Problem},
  author={Mordacchini, Matteo and Carlini, Emanuele and Dazzi, Patrizio},
  booktitle={Proc. of the 45th {IEEE International Conference on Distributed Computing Systems (ICDCS2025)}},
  year={2025}
}

@article{jiang2022graph,
  title={Graph-based deep learning for communication networks: A survey},
  author={Jiang, Weiwei},
  journal={Computer Communications},
  volume={185},
  pages={40--54},
  year={2022},
  publisher={Elsevier}
}

@article{hetzel2021graph,
  title={Graph representation learning for single-cell biology},
  author={Hetzel, Leon and Fischer, David S and G{\"u}nnemann, Stephan and Theis, Fabian J},
  journal={Current Opinion in Systems Biology},
  volume={28},
  pages={100347},
  year={2021},
  publisher={Elsevier}
}

@article{xue2025data,
  title={Data Science in Transportation Networks with Graph Neural Networks: A Review and Outlook},
  author={Xue, Jiawei and Tan, Ruichen and Ma, Jianzhu and Ukkusuri, Satish V},
  journal={Data Science for Transportation},
  volume={7},
  number={2},
  pages={10},
  year={2025},
  publisher={Springer}
}

@article{CARLINI2026108167,
title = {Dynamic workload balancing in decentralized edge systems: A marginal cost approach},
journal = {Future Generation Computer Systems},
volume = {176},
pages = {108167},
year = {2026},
issn = {0167-739X},
author = {Emanuele Carlini and Patrizio Dazzi and Luca Ferrucci and Jacopo Massa and Matteo Mordacchini},
}

@misc{Rucci:2025Asonam,
      title={Decentralized and Self-adaptive Core Maintenance on Temporal Graphs}, 
      author={Davide Rucci and Emanuele Carlini and Patrizio Dazzi and Hanna Kavalionak and Matteo Mordacchini},
      year={2025},
      eprint={2510.00758},
      archivePrefix={arXiv},
      primaryClass={cs.DC},
      url={https://arxiv.org/abs/2510.00758}, 
}

\end{document}